\documentclass[runningheads]{svmult}

\usepackage{makeidx}   % allows index generation
\usepackage{graphicx}  % standard LaTeX graphics tool
\usepackage{subeqnar}  % subnumbers individual equations
\usepackage{multicol}  % used for the two-column index
\usepackage{physprbb}  % modified textarea for proceedings,
\makeindex             % used for the subject index
\begin{document}
\title*{Synthesis Effects on the Magnetic and\protect\newline
Superconducting Properties of RuSr$_{2}$GdCu$_{2}$O$_{8}$}
\toctitle{Synthesis Effects on the Magnetic and
\protect\newline Superconducting Properties of RuSr$_{2}$GdCu$_{2}$O$_{8}$}
% allows explicit linebreak for the table of content
%
%
\titlerunning{Synthesis effects on magnetic and superconducting properties
of Ru-1212}
% allows abbreviation of title, if the full title is too long
% to fit in the running head
%
%
\author{Roberto Masini\inst{1}
\and Cristina Artini\inst{2}
\and Maria Roberta Cimberle\inst{3}
\and Giorgio Andrea Costa\inst{2}
\and Marilena Carnasciali\inst{2}
\and Maurizio Ferretti\inst{2}}
\authorrunning{Roberto Masini et al.}
% if there are more than two authors,
% please abbreviate author list for running head
%
%
\institute{CNR -- IENI, Sezione di Milano, Via Cozzi 53, 20125, Milano, Italy
\and INFM and DCCI, University of Genoa, Via Dodecaneso 31, 16146 Genova, Italy
\and CNR -- IMEM, Sezione di Genova, Via Dodecaneso 33, 16146 Genova, Italy}

\maketitle              % typesets the title of the contribution

\begin{abstract}
A systematic study on the synthesis of the Ru-1212 compound by preparing a series of samples that were annealed at
increasing temperatures and then quenched has been performed. It results that
the optimal temperature for the annealing lies around 1060-1065\r{}C; a further temperature increase worsens the
phase formation. Structural order is very important and the subsequent grinding and annealing
improves it. Even if from the structural point of view the samples appear
substantially similar, the physical characterizations highlight great
differences both in electrical and magnetic properties related to intrinsic
properties of the phase as well as to the connection between the grains
as inferred from the resistive and the Curie Weiss behaviour at
high temperature as well as in the visibility of ZFC and FC magnetic signals.
\end{abstract}

\section{Introduction}
There have been a number of reports on the coexistence of magnetic order and
superconductivity in the ruthenocuprate RuSr$_{{\rm 2}}$GdCu$_{{\rm
2}}$O$_{{\rm 8}}$, synthesized for the first time in 1995 [1]. Its
peculiarity lies in the fact that, unlike previous compounds, magnetic order
occurs at a temperature much higher than the superconducting transition
temperature. This compound is characterised by a triple perovskitic cell
similar to the high temperature superconducting cuprate (HTSC) YBa$_{{\rm
2}}$Cu$_{{\rm 3}}$O$_{{\rm x}}$, in that it contains two CuO$_{{\rm 2}}$
layers while the CuO chains are replaced by a RuO$_{{\rm 2}}$ layer.
However, various experimental reports came to different conclusions. It has
been suggested on the basis of transport measurements that its electronic
behaviour is similar to an underdoped HTSC [2] while, on the contrary, NMR
measurements resulted comparable to those of an optimally doped HTSC [3].
Some other reports concluded that the magnetic order is ferromagnetic in the
RuO$_{{\rm 2}}$ layers [2,4,5,6] in which case there should be competition
between the superconducting and magnetic order parameters resulting
eventually in a spontaneous vortex phase formation or spatial modulation of
the respective order parameters. However, powder neutron diffraction showed
that the low-field magnetic order is predominantely antiferromagnetic [7],
with a small ferromagnetic component presumibly produced by spin canting.
The spectrum of published data includes also non superconducting samples
showing similar macroscopic magnetic behaviour [8], samples showing zero
resistance but no diamagnetic signal and finally samples with evidence of a
resistive and magnetic transition. Since the physical properties of this
rutheno-cuprate material are strongly dependent on the details of the
preparation procedure, and can be very different even in samples that turn
out to be formally identical to a standard structural and chemical-physical
characterization, we have conducted a systematic on the effects of sample
preparation conditions on the properties of such hybrid compound.

\section{Experimental}
The crystal structure was determined by powder
X-ray diffraction (XRD) using Cu K$_\alpha $ radiation. Dc resistivity and
magnetic measurements were performed by the standard four-probe technique
with 1 mA current in a closed-cycle helium cryostat in the temperature range
15 - 300 K and by a Quantum Design SQUID magnetometer respectively.
Measurements were performed on similar size bar-shaped sintered
polycrystalline specimens allowing comparison of the results.

\subsection{Sample preparation}
Polycrystalline samples with nominal composition RuSr$_{{\rm 2}}$GdCu$_{{\rm
2}}$O$_{{\rm 8}}$ (hereafter referred as Ru-1212) are commonly prepared by
solid-state reaction technique from a mixture of high purity RuO$_{{\rm
2}{\rm} }$(99.95\% ), Gd$_{{\rm 2}}$O$_{{\rm 3}{\rm} }$(99.99\% ), CuO
(99.9\% ) and SrCO$_{{\rm 3}}$ (99.99\% ) [1,4,5,9,10,11]. The raw materials
are:

\noindent i -- first reacted in air at about 960\r{}C to decompose SrCO$_{{\rm 3}}$,

\noindent ii -- heated in flowing N$_{{\rm 2}}$ at 1010\r{}C,

\noindent iii -- annealed in flowing O$_{{\rm 2}}$ at temperatures ranging from 1050 to 1060\r{}C and

\noindent iv - finally, a prolonged anneal in flowing O$_{{\rm 2}}$ at 1060\r{}C is performed,
during which the material densifies, granularity is substantially reduced
[12] and ordering within the crystal structure develops [13].

\noindent Because the superconducting and magnetic
properties are affected by the details of the preparation process, which in
turn affect the microscopic structure, a systematic work on the synthesis of
Ru-1212 and the effects of sample preparation on the magnetic and
superconducting properties was developed [14]. Basically a procedure as
described commonly in literature and sketched in Fig.1 has been adopted with
the aim to give insight on the formation and stability of the various phases
involved in the synthesis of this complex system.
Each reaction step was
carried out on a MgO single crystal substrate to prevent reaction with the
alumina crucible. Between each step the products were throughly ground and
pressed into pellets.

\begin{figure}[ht]
\centering\noindent
%\begin{center}
\includegraphics[width=.6\textwidth]{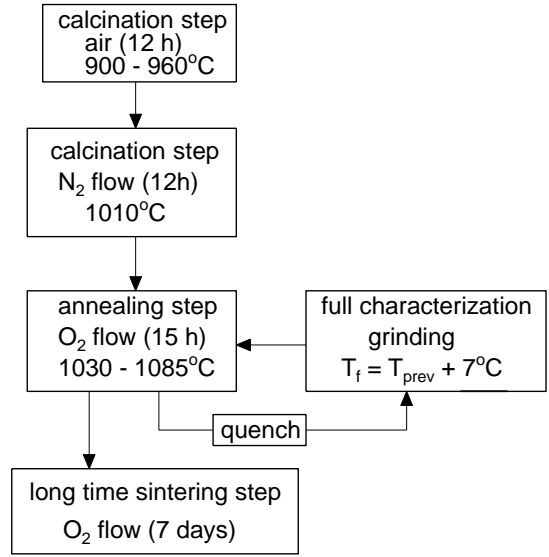}
%\end{center}
\caption[]{Sample synthesis and sintering of RuSr$_{{\rm 2}}$GdCu$_{{\rm
2}}$O$_{{\rm 8}}$}
%\label{eps1}
\end{figure}

\bigskip
\noindent
i - The stoichiometric oxides were first calcined in air at different
temperatures, T$_{{\rm A}}$, for 12 h. XRD spectra performed on these
calcined samples are shown in Fig.2 (a), (b), (c). The spectra show the peaks
of the Ru-1212 phase whose amount increases with the temperature of the
thermal treatment. There are however reflections of second phases identified
as SrRuO$_{{\rm 3}}$ and Gd$_{{\rm 2}}$CuO$_{{\rm 4}}$, with higher amounts
in samples calcined at the lower temperature, diminuishing with increasing
the calcination temperature.

\begin{figure}[ht]
\centering\noindent
%\begin{center}
\includegraphics[width=.7\textwidth]{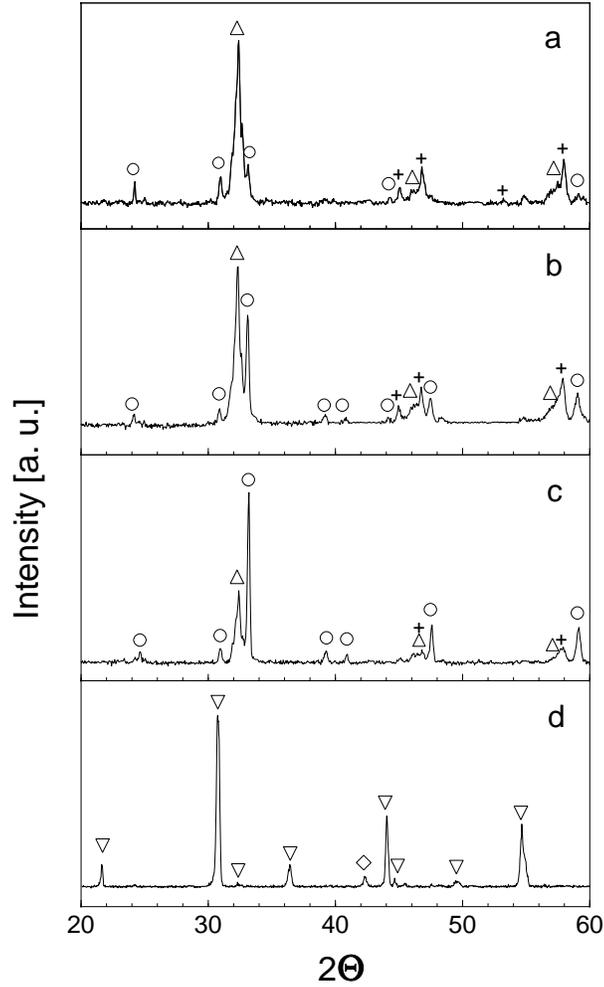}
%\end{center}
\caption[]{X-ray spectra for samples calcined at (\textbf{a}) T$_{{\rm A}}$ = 900,
(\textbf{b}) 940 and (\textbf{c}) 960\r{}C in air for 12h. $\bigcirc$ Ru--1212 , $\bigtriangleup$ SrRuO$_{{\rm 3}}$
and + Gd$_{{\rm 2}}$CuO$_{{\rm 4}}$; (\textbf{d}) after annealing in N$_{{\rm 2}}$
for 15h  $\bigtriangledown$  Sr$_{{\rm 2}}$GdRuO$_{{\rm 6}}$, $\diamond$
 Cu$_{{\rm 2}}$O}
%\label{eps2}
\end{figure}

\noindent
ii - The pellets were then annealed in flowing nitrogen at 1010\r{}C for
12 h. The sintering in N$_{{\rm 2}}$ gas is required to suppress the
SrRuO$_{{\rm 3}{\rm} }$ phase [10]. This step resulted in fact in the
formation of a mixture of Sr$_{{\rm 2}}$GdRuO$_{{\rm 6}{\rm} }$ and Cu$_{{\rm
2}}$O independently of the starting calcined mixture from step (i)
(hereafter named L-serie). Typical XRD pattern is shown in Fig.2 (d)
obtained from the sample calcined at 900\r{}C which contained the highest
amount of SrRuO$_{{\rm 3}}$. No detectable traces, within the resolution of
the technique, of such very stable in oxidising environment [10] impurity phase,
were observed.

On behalf of these results, the synthesis of Ru-1212 by using SrO$_{{\rm
2}}$ as starting reagent in place of SrCO$_{{\rm 3}}$ was investigated
(sample I). Raw materials were then heated directly in N$_{{\rm 2}}$ flow at
1010\r{}C avoiding thus the first calcination step (i) in air. No
significant differences were obtained in the composition of the products as
inferred from XRD analysis with respect to previous results shown in Fig.2(d).

\bigskip

\noindent
iii -- The L- serie mixture was then subjected to eight successive sintering
steps in flowing O$_{{\rm 2}}$, each one lasting 15 h, at successively
increasing temperatures in the range 1030\r{}C -- 1085\r{}C. Each
successive thermal treatment was performed at a temperature about 7\r{}C
higher than the previous one. In order to investigate the effects of the
thermal treatments the product was quenched to room temperature at the end
of every step, fully characterized, reground, pressed into pellets and
subjected to the successive thermal treatment.

Powder XRD patterns of all our samples show Ru-1212 as the major phase, with
zero to some amount of SrRuO$_{{\rm 3}{\rm} }$ as minor impurity depending
on the sample preparation condition. Traces of second phase SrRuO$_{{\rm 3}}$
(2\% vol. for sample L1) with decreasing amount up to sample
L3 were detected. Single phase materials were obtained afterwards. All peaks can be indexed assuming a tetragonal
lattice and Table 1 lists the lattice parameters calculated for these
Ru-1212 samples. In Fig.3 the X-ray powder diffraction pattern of sample L5,
synthesized after five sintering steps up to 1067\r{}C for a total time t=
(15 x 5) = 75 h, is reported.

\begin{figure}[ht]
\centering\noindent
%\begin{center}
\includegraphics[width=.65\textwidth]{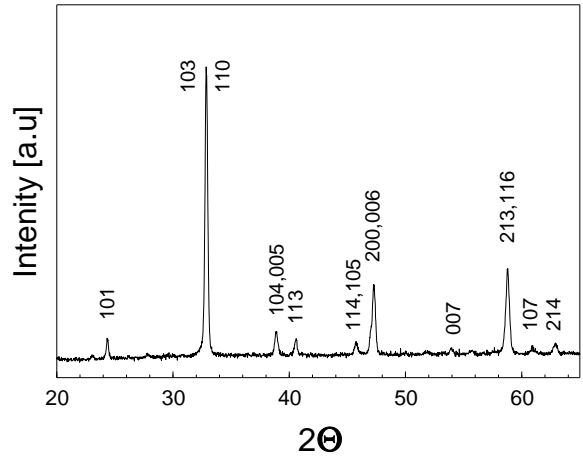}
%\end{center}
\caption[]{XRD pattern of sample L5}
%\label{eps3}
\end{figure}

The same XRD spectra have been obtained for sample I, subjected to a
subsequent thermal treatment at 1050\r{}C for 24 h and successively to a
prolonged anneal at 1060\r{}C for a week in O$_{{\rm 2}}$ flow.

\begin{table}
\caption{Synthesis, structural and electrical data of L--serie Ru--1212
samples}
\begin{center}
\renewcommand{\arraystretch}{1.4}
\setlength\tabcolsep{5pt}
\begin{tabular}{lllllll}
\hline\noalign{\smallskip}
sample & $T_{ann}(^\circ$C) & a ($\AA$) & c ($\AA$) & $\varrho_{290}(m\Omega$cm) & $T_{R=0}$(K) & $T_{max}$(K) \\
\noalign{\smallskip}
\hline
\noalign{\smallskip}
L1 & 1031 & 3.826(2) & 11.516(7) & 154.8 & --$^{\mathrm a}$ & 37 \\
L2 & 1037 & 3.821(3) & 11.528(8) & 141.7 & --$^{\mathrm a}$ & 44 \\
L3 & 1044 & 3.831(1) & 11.545(7) & 140.9 & --$^{\mathrm a}$ & 46 \\
L4 & 1053 & 3.831(1) & 11.547(6) & 70.6 & 17 & 48 \\
L5 & 1061 & 3.828(1) & 11.552(6) & 30.6 & 25 & 49 \\
L6 & 1067 & 3.844(1) & 11.585(3) & 14.0 & 21 & 44 \\
L7 & 1073 & 3.845(1) & 11.610(5) & 23.7 & 21 & 47 \\
L8 & 1084 & 3.835(1) & 11.580(5) & 22.0 & --$^{\mathrm a}$ & 45 \\
\noalign{\smallskip}
\hline
\noalign{\smallskip}
\end{tabular}
\end{center}
$^{\mathrm a}$ No information available below 15 K. See text for a complete discussion. \\
\label{Tab1b}
\end{table}

Parallel checks have been performed allowing us to conclude that
reaching the ``optimal'' temperature directly in one step for a time which
is the sum of the corresponding partial times of each single step covered up
to the same temperature does not produce the same results of the longer
procedure described above. Single-phase formation seems to be kinetically hindered by the slow
decomposition rate of the impurities which already formed upon calcination.
Repeated homogenisations, related to the sequence of grinding and annealing,
improve the phase purity of the material and
control the superconducting behaviour.

Morphologically, all the L-series samples show a high grain homogeneity with
clean grain boundaries as probed by SEM and microprobe analyses. Fig.4
shows the typical granular morphology detectable at the beginning of the
thermal treatment cycles (sample L2) with an average grain size of about 2
$\mu $m.

\begin{figure}[h]
\centering\noindent
\includegraphics[width=.65\textwidth]{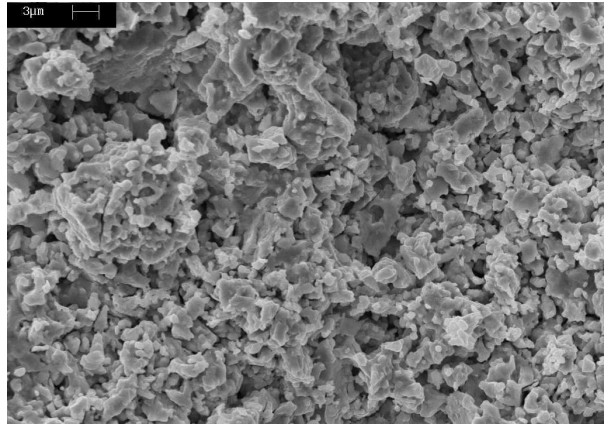}
\caption[]{SEM picture of sample L2}
\end{figure}

A progressive grain growth and a corresponding increase in grain
connectivity due to the different thermal treatments can be observed (Fig.5,
sample L6, it can be noticed how some grains begin to coalesce
into big aggregates dispersed in an almost unchanged granular matrix),
without reaching, by the way, complete sintering at the highest temperature.

\begin{figure}[h]
\centering\noindent
%\begin{center}
\includegraphics[width=.65\textwidth]{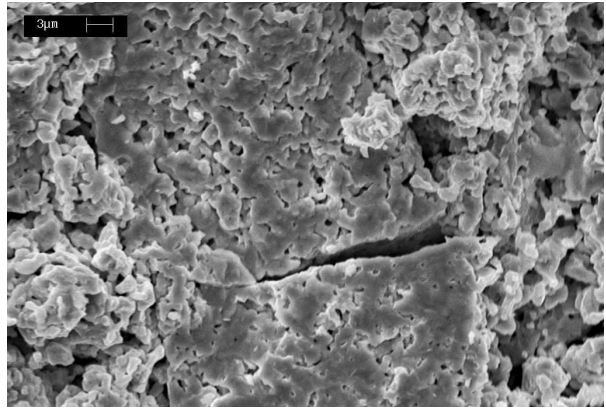}
%\end{center}
\caption[]{SEM picture of sample L6}
%\label{fig5}
\end{figure}

Such behaviour is related to the difference between
the decomposition temperature of the 1212 phase and the maximum temperature
of the thermal processes considered in this work.

A geometric density variation of about 5\% between the first and the final
bulk sample has been measured (d $ \approx $ 4.2 g/cm$^{{\rm 3}}$
corresponding to about 63\% of the theoretical crystallographic density).

\section{Electrical properties}
In general resistivity measurements of Ru-1212 show a superconducting
transition at 45 K with a very slight upturn in the vicinity of T$_{{\rm
c}}$ reaching zero resistivity at a lower temperature between 20 -- 30 K.
The resistivity transitions at H = 0 are much broader than those observed in
many of the other HTSC. A metallic behaviour, with a T linear dependence at
high temperatures above 100 K, is usually observed. A magnetic transition at
T$_{{\rm M}}$ = 132 K manifest itself as a small yet noticeable kink/minimum
in the resistivity related to the onset of the
magnetic ordering of the Ru lattice.

Since the oxygen stoichiometry is pratically
unchanged in Ru-1212, the annealing turns out to influence mainly the
granularity and ordering within the crystal structure. Previous studies
have shown that the semiconductor-like upturn and the zero resistivity
temperature are critically dependent on the sample processing. [10,15]. In
particular, according to [6] the slight upturn in the vicinity of T$_{{\rm
c}}$ is related to grain boundary effects. High resolution TEM study on
Ru-1212 has shown that prolonged thermal treatment at 1060\r{}C in
O$_{{\rm 2}}$ removes most of a multidomain structure, consisting
predominantly of 90\r{}  rotations, as well as significantly reduces the
semiconductor-like upturn [15]. Part of the superconducting transition width
may be due to structural disorder. However it must be underlined that a
broad superconducting transition is also expected within the spontaneous
vortex phase model [16].

\begin{figure}[ht]
\centering\noindent
%\begin{center}
\includegraphics[width=.7\textwidth]{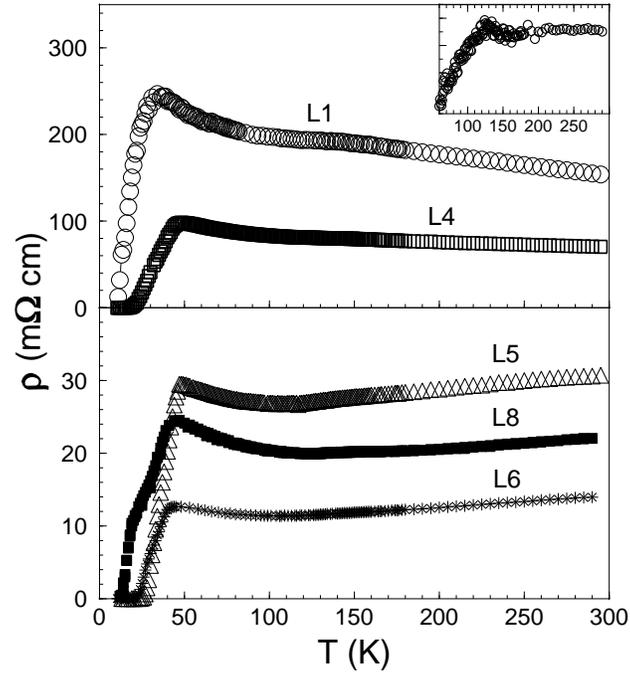}
%\end{center}
\caption[]{Resistivity vs T curves of some selected L-serie samples synthesized
under different conditions. Inset: d$\varrho$/dT temperature dependence for
sample L4 around $T_{M}$ }
%\label{eps6}
\end{figure}

Curves of $\varrho $T) of selected samples (L-serie) considered significative,
for sake of clarity, of the overall process of synthesis are shown in Fig.6.
All samples exhibit weakly pronounced or local minima in the dc
resistivity near the magnetic transition temperature, of the order of about
132 K. This feature is more clearly visible in the inset of the figure where
the derivative of the resistivity (sample L4) is plotted.

At low temperatures the dc resistance shows a semiconductor-like upturn
followed by a sudden decrease in resistivity starting at T$_{{\rm m}{\rm
a}{\rm x}}$ and achieving zero resistivity state for temperatures below 30 K
as reported in Table 1. There is a small increase in the zero resistivity
temperature for our best sample (L6) and only a small reduction in the
semiconductor-like upturn. Summarizing the general trend, it can be stated
that the resistivity is progressively decreased and a crossover from a
semiconducting to metallic normal state resistivity behaviour is observed on
going from L1 to L8 sample. We underline that zero resistivity has not been
reached for samples from L1, L2, L3 and L8 even if, considering their strong
resistivity drop detected below 45 K, a R = 0 value is expected at a
temperature lower than 13 K for samples L2, L3 and L8.
A comparison between the resistivity behaviours, independently of their
granular nature, has been possible because different values are not related
to the sample density variations, which as already noted is almost unchanged
for all the samples.

\begin{figure}[h]
\centering\noindent
%\begin{center}
\includegraphics[width=.5\textwidth]{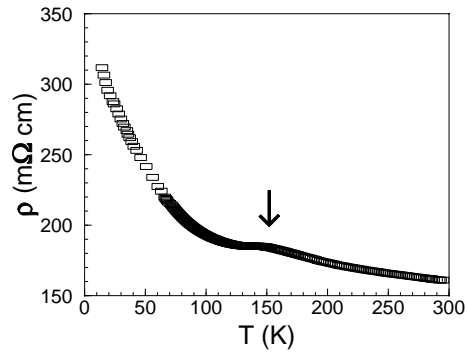}
%\end{center}
\caption[]{Resistivity temperature dependence of sample I}
%\label{eps7}
\end{figure}

A semiconducting-like transport with no indications of transition to
superconductivity at low temperatures is observed for sample E in all the
temperature range considered, as shown in Fig. 7. It is noteworthy that a
kink in resistivity is observed in the vicinity of T$_{{\rm M}}$ (arrow in
Fig. 7). This anomaly is due to a reduction of spin scattering. Such a
behaviour is also observed in SrRuO$_{{\rm 3}}$ single crystals [17] at
around its ferromagnetic transition temperature.

The derivative of resistivity, shown in Fig. 8 for L3 -- L8 samples, clearly
shows two overlapping maxima, indicating that the resistive transition
proceeds in two steps: a high temperature contribution, associated with the
thermodynamic superconducting transition temperature and another one, at a
lower temperature, which critically depends on sample processing conditions
[10,14,15] as well as the zero resistance temperature value.

\begin{figure}[ht]
\centering\noindent
%\begin{center}
\includegraphics[width=.5\textwidth]{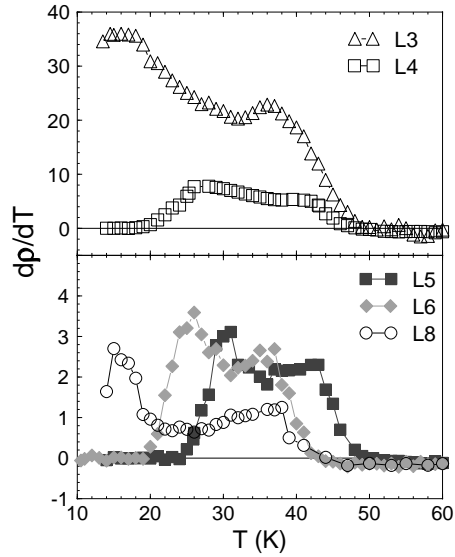}
%\end{center}
\caption[]{d$\varrho$/dT temperature dependence}
%\label{eps8}
\end{figure}

Since from the analysis of x-ray patterns our superconducting and non
superconducting samples are indistinguishable, more insight about the
physical nature of the superconducting and magnetic states is expected from
magnetic measurements.

\section{Magnetic properties}
The magnetic characterization of the ruthenocuprate materials is a crucial
and not trivial point. Magnetic measurements are obviously a key tool to
observe both the superconducting and magnetic behaviour of these samples,
but many years after their successful synthesis [1] and in spite of a
great experimental effort devoted to this problem, many doubts still survive
about the magnetic ordering present in these type of samples [7, 11, 18, 19,
20, 21, 22].

First of all, we recall that generally granular samples are measured:
therefore, all the problems related to the granular behaviour of HTSC and,
in general, to the distinction between intrinsic and extrinsic properties
(intra-granular and inter-granular) must be born in mind.

A first problem encountered in the observation of the superconducting
behaviour is the fact that the standard diamagnetic signals, both in the
Field Cooled (FC) and Zero Field Cooled (ZFC) mode, are not always seen in
all the samples of such compounds [13, 14, 18, 23]: what is more often
observed is the shielding signal, rarely the diamagnetism related to the FC
procedure. Both signals are quickly removed by the application of even a
small external magnetic field (few tens of Gauss). In contrast, even when in
the magnetic measurement there is no trace of superconducting behaviour, it
may be observed resistively and the application of even a high external
magnetic field (up to Tesla) does not destroy it [24]. The reason for such
a contradictory phenomenology may be understood bearing in mind the
simultaneous presence of magnetic and superconducting ordering in these
samples. This fact implies consequences both on the sample physical
behaviour and on the measurement technique used to monitor it. We recall
that $\mu $SR measurements [5] indicate the homogeneous presence of an
internal field that, at low temperature, may reach hundreds of Gauss and may
give rise to a spontaneous vortex phase (SVP) in the temperature range where
it exceeds the first critical field H$_{{\rm c}{\rm 1}}$(T) [25]. In a type
II superconductor at H$>$H$_{{\rm c}{\rm 1}}$ the Meissner effect is
practically never observed for the presence inside the materials of
``pinning centres'' that are able to block the flux lines and prevent their
expulsion. This is the reason why the FC diamagnetic signal may be very
small and its difference from the ZFC signal is an indication of the
critical current density that a sample can carry. A vast literature related
to high T$_{{\rm c}}$ superconductors illustrates unambiguously this item
[26, 27]. Moreover, as noted in [28], the magnetization of ruthenocuprate
materials contains magnetic signals arising from different contributions:
the Gd paramagnetic spin lattice, the Ru spin lattice and, finally, the
diamagnetic signal related to the superconducting behaviour. Both for Gd and
Ru spin lattice the antiferromagnetic ordering is coupled with a
ferromagnetic component that, in the case of Ru, is attributed to a canting
of the lattice and in the case of Gd is simply related to the presence of
the net ferromagnetic moment of the Ru lattice [20]. The simultaneous
presence of such opposite magnetic signals makes the magnetic measurement
unsuitable for the observation of the superconductivity: in fact, such
measurement cannot separate the magnetic signal related to superconductivity
from that related to the magnetic ordering. Moreover, it is clear that the
application of an external magnetic field exalts the magnetic signals and
depresses the superconducting one, destroying very quickly the visibility of
the superconductivity. In the light of these considerations we can
understand the fact that the superconducting behaviour is often observed
resistively but not magnetically: it depends on the competition between two
opposite magnetic signals, one related to the magnetic ordering, the other
to the superconducting one. ``More superconductivity'' is obviously related
to many factors: the amount of superconducting phase inside the sample, the
quality of the connection between the grain that makes the shielded volume
and therefore the related magnetic signal smaller or larger, and the
intrinsic properties of the Ru-1212 phase that, as we will see, may change
in connection with the grade of order of the material.
Now, dealing with the experimental problems, we point out the following.
In order to enhance the superconducting behaviour it is suitable to apply
magnetic fields as small as possible.\textbf{} This fact, due the peculiar
modalities of elaboration of the instrumentation commonly used, must be
considered in detail. The first problem is the exact knowledge of the field
that is effectively seen by the sample, and the second is strictly related
to the complexity of the magnetic signal present in these samples. A small
remanent magnetic field in the superconducting coil of the experimental
set-up is often present. It may be zeroed by a procedure that, starting from
a value of some Tesla, applies coercive fields of decreasing intensity. In
such a way the field is zeroed but for a few Gauss that may be zeroed in the
central point of the magnet by applying a small counterfield. Anyway, a very
small residual field survives and turns out to be of the order of fractions
of Gauss. In the light of what has been said, a real ZFC measurement cannot
be made and, since the FC magnetic moment is about one order of magnitude
greater than the ZFC, also a residual field of fractions of Gauss may give a
considerable magnetic signal whose polarity depends on the field polarity.
In addition, the basic condition of a homogeneous magnetic moment required
by the SQUID magnetometer is not fulfilled, in particular at low
temperatures, where, as a consequence of the applied field, magnetic moments
of opposite polarity will be present in the sample. Finally, we recall that
during the measurements the sample is moved for a length that is usually of
few centimetres, so that it travels in a non uniform magnetic field that
makes it follow a minor hysteresis loop. If the value of the moment is not
constant during the scan, an asymmetric scan wave form will be observed and
the quality of the measurement will drastically degrade [29].

All we have said is illustrated in Fig. 9 where magnetization
measurements are reported for both ZFC and FC conditions. For sake of
clarity we report data for some representative samples only. The cuspid at T
$ \cong $ 30 K marks the magnetic ordering: there is a small variation in
this temperature, which is smaller in the sample with higher superconducting
temperature in agreement with the literature data [8]. It is remarkable to
observe the different behaviour exhibited by the various samples: L3 gives
no hint of superconductivity, L5 exhibits a very clear shielding
corresponding to about 75\% of the maximum diamagnetic signal at $\mu
$H$_{{\rm e}{\rm x}{\rm t}}$ = 0.5 G while at $\mu $H$_{{\rm e}{\rm
x}{\rm t}}$ = 5 G its transition is strongly worsened, L6 shows a
diamagnetic shift after an ascent of the magnetization (probably due to the
instrumental effects we outlined before, for the presence of two opposite
magnetic signals of similar magnitude), and L8 shows a behaviour very
similar to L3. At the lowest temperatures a large contribution from Gd
sublattice, which orders antiferromagnetically at 2.5 K, is clearly visible
in FC magnetization curves for the magnetically non superconducting samples
L3 and L8. If the superconductivity is marked by the visibility of a
diamagnetic shift of the ZFC or FC signals, such behaviour is surely absent
in L1, L2, L3, L4 and L8 while, to different extent, it is observed in L5,
L6 and L7. In the resistivity measurements, on the contrary, all the samples
show a large drop of resistivity, but at the temperature of T=13 K (the
minimum value at which we measured resistively, while magnetically we
reached T=5 K) zero is reached for L4, L5, L6 and L7 samples.

\begin{figure}[h]
\centering\noindent
%\begin{center}
\includegraphics[width=.95\textwidth]{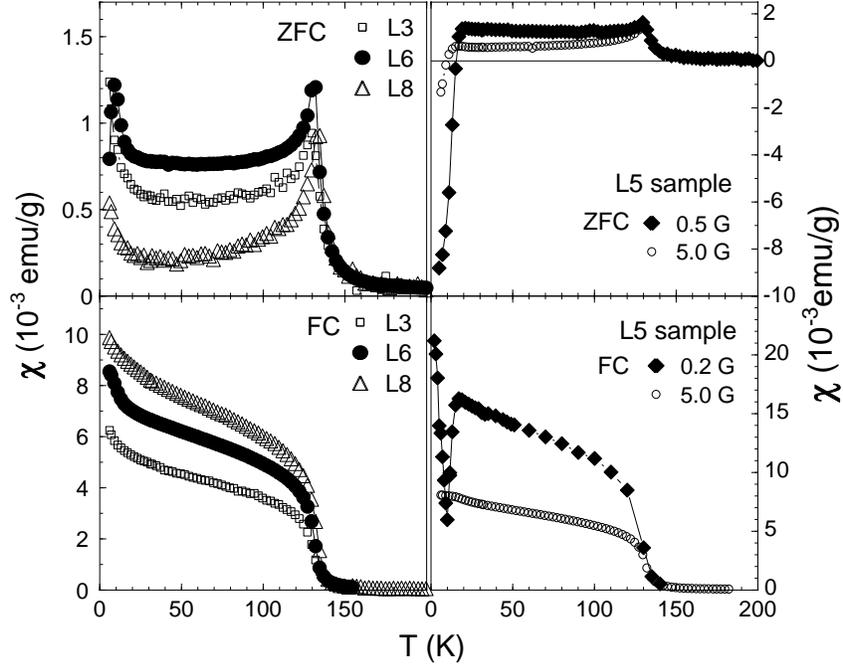}
%\end{center}
\caption[]{ZFC and FC susceptibility vs temperature curves for some samples
of the L--serie: {\it left\,} L3 (5.5 G), L6 (5.5 G) and L8 (3.0 G); {\it right\,} L5 sample}
%\label{eps9}
\end{figure}

In lower Fig. 9 the FC data for samples L3, L5, L6 and L8 are shown. A sudden
onset of a spontaneous magnetic moment appears, related to a ferromagnetic
component arising from Ru spin ordering in RuO$_{{\rm 2}}$ planes. Such a
spontaneous magnetization develops at a temperature in the 130-135 K range
and below 110 K it rises almost linearly as temperature decreases down to
about 50 K. We remark the very similar behaviour of L3 and L8, already
observed in upper part of Fig. 9. A clear diamagnetic behaviour is seen only in L5: at
the minimum applied field of 0.2 G, and to a minimum extent\textbf{\textit{
}}even at 1.2 G, a diamagnetic behaviour that quickly reenters is seen in
the FC curve. At 5.5 G the diamagnetic effect is only seen as a constant
value hindering the Gd magnetic ordering. Such behaviour has been already
observed [30].

\begin{figure}[ht]
\centering\noindent
%\begin{center}
\includegraphics[width=.8\textwidth]{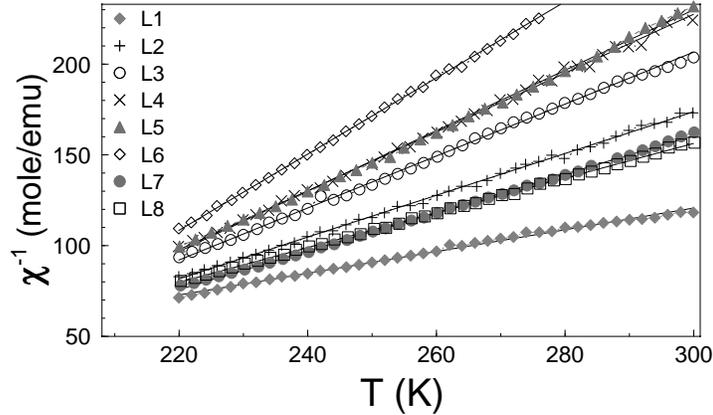}
%\end{center}
\caption[]{$\chi_{\mathrm{Ru}}^{\mathrm -1}$ vs temperature for all
the samples. Fitting results are shown as solid lines  }
%\label{eps10}
\end{figure}

In Fig. 10 we present the inverse of ruthenium susceptibility as a function
of temperature for all the samples in the series L1-L8. In the calculation
of the ruthenium susceptibility we have followed the procedure suggested by
Butera et al. [22]. Such a procedure calculates the Ru susceptibility by
subtracting three magnetic contributions to the experimental value: 1) the
paramagnetic contribution from Gd ions, 2) the core diamagnetism for the
1212 compounds as deduced from the Landolt-B\"{o}rnstein tables, and 3) a
temperature-independent Pauli-like contribution coming from the conduction
electrons. The so obtained ruthenium susceptibility is fitted by the
Curie-Weiss relationship $\chi _{Ru} = {\frac{{C_{Ru}} }{{(T - \Theta )}}}$
and allows to calculate both the Curie temperature $\Theta $ and the effective
magnetic moment $\mu _{\mathrm {eff}}$ for Ru atom.
Although a maximum content of about 2 vol.\% of SrRuO$_{{\rm 3}}$
impurity phase was detected from x--ray analyses in sample L1, with decreasing
amount to zero for sample L4, a similar negligible error on the absolute values of
$\mu_{\mathrm{eff}}^{\mathrm Ru}$ and $\theta $ has been calculated with no
significant effect on their general behaviour.

\begin{figure}[h]
\centering\noindent
%\begin{center}
\includegraphics[width=.65\textwidth]{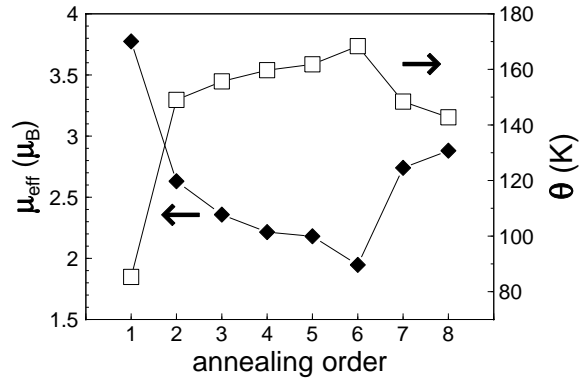}
%\end{center}
\caption[]{$\mu_{\mathrm{eff}}^{\mathrm Ru}$ and $\theta $ calculated from
a best fit of $\chi_{\mathrm{Ru}}^{\mathrm -1}$ vs temperature data}
%\label{eps11}
\end{figure}

The obtained results
are reported in Fig. 11 and in Table 2 as a function of the number of
annealing steps that give rise to the sequence L1-L8.
\begin{table}
\caption{$\mu_{\mathrm{eff}}^{\mathrm Ru}$ and $\theta$ values as a function of annnealing
steps and ``superconductivity'' status of all L--serie samples}
\begin{center}
\renewcommand{\arraystretch}{1.4}
\setlength\tabcolsep{5pt}
\begin{tabular}{llllll}
\hline\noalign{\smallskip}
sample & $\mu_{\mathrm{eff}}^{\mathrm Ru}$($\mu_{B}$) & $\theta$(K) & res$^{\mathrm a}$
& m$_{ZFC}^{\mathrm a}$ & m$_{FC}^{\mathrm a}$ \\
\noalign{\smallskip}
\hline
\noalign{\smallskip}
 L1 & 3.77 & 85.3 & --$^{\mathrm b}$ & no & no \\
 L2 & 2.63 & 149.0 & --$^{\mathrm b}$ & no & no \\
 L3 & 2.36 & 155.6 & --$^{\mathrm b}$ & no & no \\
 L4 & 2.21 & 159.7 & yes & no & no \\
 L5 & 2.18 & 161.8 & yes & yes & yes \\
 L6 & 1.95 & 168.3 & yes & yes & yes \\
 L7 & 2.74 & 148.4 & yes & yes & no \\
 L8 & 2.88 & 142.7 & --$^{\mathrm b}$ & no & no \\
\hline
\end{tabular}
\end{center}
$^{\mathrm a}$ measurement technique utilized to detect superconductivity:
resistivity, ZFC and FC magnetization. \\
$^{\mathrm b}$ No information available below 15 K. See text for a complete
discussion.
\label{Tab1a}
\end{table}

Starting from L1 the
$\Theta $ values increase, reach a maximum (around L5-L6 of about 160 K), and
then decrease going up to L8. The $\mu _{\mathrm {eff}}$ values have a
specular trend, decreasing from the value 3 for L1 down to a minimum value
of about 2 at L6, and then slightly increasing once again.
 Since the
superconductivity is better observed in the samples L5, L6, L7 both by
resistivity and magnetic measurements, these data suggest that an improved
superconducting behaviour may be related to small intrinsic variations in
the structure of the sample that produces smaller effective magnetic moments
for Ru atom and higher Curie temperatures. We give here only some
suggestions to be explored. The $\mu _{\mathrm {eff}}$ values derived by
the best fit imply that Ru is in a mixed valence state between Ru$^{{\rm
4}{\rm +} }$ and Ru$^{{\rm 5}{\rm +} }$. Such a result has been firstly
proposed by Liu et al. [21] through XANES spectroscopy and successively
confirmed by Butera et al. [22] through magnetic measurements by means of
the procedure we have outlined. These results definitively contradict the
hypothesis that Ru exhibits an effective moment $\mu _{\mathrm {eff}} \cong$
1 $\mu_{B}$/Ru atom, as proposed in [4]. On passing from L1 up to L8
the proportion of Ru$^{{\rm 4}{\rm +} }$ and Ru$^{{\rm 5}{\rm +} }$ changes.
Possible consequences of this fact are: slight variations in the carriers
number and, as a consequence, in the critical temperature (see the
resistivity data in Fig.6 and Table 1),
different coupling between the superconducting and the magnetic planes both
in term of total magnetic moment seen by the conduction electron (with
increased or decreased pair-breaking effect) and in term of coupling between
orbitals of superconducting and magnetic electrons [31]. The origin of these
variations may be found in a different degree of lattice disorder following
the various annealing steps performed at different temperatures that, as we
have widely observed, produce significant variations in all the physical
properties. Moreover, the lattice disorder can imply a certain amount of Cu$
\to $Ru substitutions that are a possible candidate for the observed
variations of the effective magnetic moment. Also the variation of $\Theta
$ may be the consequence of the different coupling between Ru atoms following
the different valence state.

\begin{figure}[h]
\centering\noindent
%\begin{center}
\includegraphics[width=.65\textwidth]{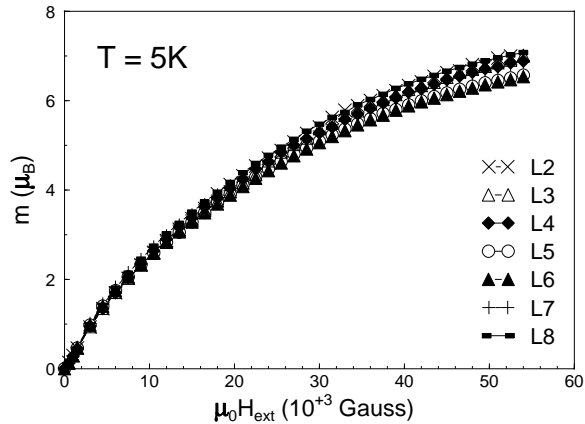}
%\end{center}
\caption[]{Magnetic moment versus magnetic field at 5K for L--series samples}
%\label{eps12}
\end{figure}

The same trend we have seen in $\mu _{\mathrm {eff}}$ is observed in the saturation moment as seen by measuring
magnetization up to the maximum field of 5.5 Tesla at T=5 K. Results are
shown in Fig 12. The values change from the minimum value of 6.5 $\mu_{B} $ for
L5 and L6 samples up to a maximum value of 7 $\mu_{B} $ in the L2 and
L8 samples. We remark that in the experimental conditions we have used the
saturation is not completely reached, but the hierarchy of the saturated
magnetic moments is surely correct.

\section{Conclusions}
The magnetic and superconducting properties of Ru--1212 have been studied
and compared for a series of samples synthesized under different conditions
with the aim to find out the fundamental parameters ruling out the phase
formation and its related structural and physical properties. From our
experimental work it results that the optimal annealing temperature lies
in a narrow temperature range around 1060-1065\r{}C; further temperature
increase worsens the phase formation. Subsequent grinding and annealing
steps up to this temperature improve the phase homogeneity.
A wide range of physical properties has been obtained on quenched samples
from the same batch, which differ only in the synthesis procedure parameters.
No other substantial differences were detected for these samples, all showing
similar compositional and structural characteristics. It emerges that the
preparation method plays an important role when dealing with the magnetic
and superconducting properties of this hybrid compound. So far, published
data on the Ru--1212 phase show the same general trend for what regards
the measured physical properties. Because most of the samples are chemically
and structurally comparable, great care must be taken in the preparation
process details such as the final sintering temperature and the number
of homogeneization steps (if any) performed up to that temperature. Only
samples with the same thermal history/parameters can be compared.

%INDEX%%%%%%%%%%%%%%%%%%%%%%%%%%%%%%%%%%%%%%%%%%%%%%%%%%%%%%%%%%%%%%%
% Please check with the editor of your book whether he plans to
% include a "mutual" subject index - if so, please code your entries
% in the standard syntax. For your own purposes you may print your
% "personal" index by using the following commands:
%
%\clearpage
%\addcontentsline{toc}{section}{Index}
%\flushbottom
%\printindex
%%%%%%%%%%%%%%%%%%%%%%%%%%%%%%%%%%%%%%%%%%%%%%%%%%%%%%%%%%%%%%%%%%%%%

\end{document}